\documentclass[aps,prd,twocolumn,superscriptaddress,groupedaddress]{revtex4}  
\usepackage{graphicx}
\usepackage{amsmath}
\usepackage{amssymb}
\usepackage{amsthm}
\usepackage{bbm}
\DeclareMathOperator{\Tr}{Tr}

\newcommand{\bra}[1]{\ensuremath{\left\langle #1 \right|}}
\newcommand{\ket}[1]{\ensuremath{\left| #1 \right\rangle}}

\renewcommand{\dag}{\dagger}

\usepackage[colorlinks=true, linkcolor=blue, urlcolor=blue, citecolor=blue, anchorcolor=blue]{hyperref}

\begin{document}
\title{Tensor renormalization group study of the non-Abelian Higgs model in two dimensions}
\author{Alexei Bazavov}
\affiliation{Department of Computational Mathematics, Science and Engineering, and Department of Physics and Astronomy, Michigan State University, East Lansing, MI 48824, USA}
\author{Simon Catterall}
\author{Raghav G. Jha}
\author{Judah Unmuth-Yockey}
\email{jfunmuthyockey@gmail.com}
\affiliation{Department of Physics, Syracuse University, Syracuse, NY 13244, USA}
\date{\today}
\begin{abstract}
    We study the $SU(2)$ gauge-Higgs model in two Euclidean dimensions using the tensor renormalization group (TRG) approach.  
    We derive a tensor formulation for this model in the unitary gauge and 
    compare the expectation values of different observables between TRG and Monte Carlo simulations finding 
    excellent agreement between the two methods. In practice we find the TRG method to be far superior to Monte
    Carlo simulation for calculations of the Polyakov loop correlation function which is used to extract the static quark potential.
\end{abstract}
\maketitle

\section{Introduction}
It is usually very difficult to extract the emergent, long distance properties of quantum field theories or many-body systems from the underlying partition function. A powerful approach pioneered by Wilson known as the real-space renormalization group attempts to replace the elementary degrees of freedom by new averaged--or block--variables at larger scales. In order to maintain the correct long distance behavior under one such blocking
requires a change in the effective coupling constants of the theory. If this procedure is applied recursively one generates a description of the theory at longer and longer length scales accompanied by a corresponding flow in the effective couplings. 

In the original scheme due to Wilson and Kadanoff, this coarse graining procedure was carried out on the original fields and their corresponding Hamiltonian or action. However, in recent years it has been appreciated that it is sometimes more efficient to carry out this operation on alternative representations of the partition function called tensor networks. Algorithms that attempt to compute the partition function (or low lying states in a Hamiltonian formulation) by a recursive  blocking of these tensors are called tensor renormalization group (TRG) methods.
In the last decade there have been many such proposals \cite{Levin:2007,Xie:2009,Xie:2012,Vidal:2015} and intriguing connections have been drawn between tensor networks such as MERA (multi-scale entanglement renormalization ansatz) \cite{Vidal:2006} which was designed to capture the behavior of critical systems and the AdS/CFT correspondence \cite{Swingle:2009, Swingle:2012}. Tensor networks have also been used to study gauge theories and their real-time dynamics \cite{PhysRevX.6.011023} . The implication of gauge symmetry in one such tensor network - the matrix product state (MPS)- applied to the Schwinger model was discussed and used to calculate the confining potential in \cite{Buyens:2013yza, Buyens:2015tea}. 

In this paper, we will derive an explicit tensor network representation of a two-dimensional non-Abelian
gauge theory coupled to matter and show how a particular TRG method - the HOTRG algorithm \cite{Xie:2012} - can be used to efficiently
calculate the free energy and other observables in the theory. We will compare these results with conventional Monte Carlo (MC) calculations to test the validity of our tensor renormalization group procedure.  A similar study was carried out for the Abelian version of this model in Refs.~\cite{abelian-higgs,abelian-ploop}, along with a study of the Schwinger Model in Ref.~\cite{grassmann-tensor}.  
Reformulations of models using similar
discrete variables offer an alternative starting point for other sampling methods such as worm algorithms, \emph{e.g.} Refs.~\cite{su2chiral-worm,worm-schwinger}

Our results indicate that the tensor methods are typically much more 
efficient than the Monte Carlo method for the computation of observables in such theories. 
Furthermore, tensor networks are also promising for studying theories with a sign problem, where the Monte Carlo methods cannot be applied. 

The paper is organized as follows: In Sec.~\ref{sec:model}, we introduce the model studied in this paper and define notations.  In Sec.~\ref{sec:tensor-con}, we outline the tensor formulation for the model.  In Sec.~\ref{sec:limits} we review some analytic limits of the model and use this to check the formulation and numerical results. In Sec.~\ref{sec:mc-compare}, we compute several observables and compare them to the our Monte Carlo results. Finally, in Sec.~\ref{sec:conclusions} we give concluding remarks and discuss future directions for this work.

\section{The Model}
\label{sec:model} 
We consider the non-Abelian gauge-Higgs model with the group $SU(2)$ in two Euclidean dimensions.  This model was studied in Ref.~\cite{Gongyo:2014jfa} using Monte Carlo methods. In the continuum the action contains both a Yang-Mills and a scalar kinetic term. For the pure Yang-Mills term we have,
\begin{equation}
\label{eq:ym-action}
    S_{g} = -\frac{\tilde{\beta}}{2} \int d^2 x \Tr \Big(F_{\mu \nu} F^{\mu \nu}\Big)
\end{equation}
with $F_{\mu \nu}$ the field strength tensor in the fundamental representation and $\tilde{\beta} = 1/g^2$ the inverse coupling. For the scalar kinetic
term we have,
\begin{equation}
    S_{\Phi} = \frac{\kappa}{4} \int d^2 x \Big(D_{\mu}\Phi\Big)^{\dagger} \cdot \Big(D_{\mu}\Phi \Big)
\end{equation}
with $D_{\mu} = \partial_{\mu} + A_{\mu}$ the covariant derivative and $\Phi$ a complex doublet scalar field.  The scalar potential for this field is given by
\begin{equation}
    V(\Phi) = \int d^{2}x \; (\Phi^{2} + \lambda (\Phi^{2} - 1)^{2}).
\end{equation}


The different terms in the continuum action map straightforwardly to their lattice analogs. Here we work on a lattice with dimensions $N_{s} \times N_{\tau}$, with $N_{s}$, $N_{\tau}$ the number of lattice sites in the spatial and temporal directions, respectively.
For the pure Yang-Mills term we use the standard Wilson action,
\begin{equation}
\label{eq:lat-ym}
    S_{g} = -\frac{\beta}{2} \sum_{x} \Tr \left[U_{x,1} U_{x+\hat{1},2} U^{\dagger}_{x+\hat{2},1} U^{\dagger}_{x,2}\right]
\end{equation}
where one takes a product of the gauge fields associated with the links around an elementary square (plaquette) for each square of the lattice.  Each link variable is defined as $U_{x,\mu} = e^{-a A_{x,\mu}}$ with $a$ the lattice spacing, $A_{x,\mu} = -i g A_{x,\mu}^{i}T^{i}$ the vector potential with the $SU(2)$ generators in the fundamental representation, and $x$, $\mu = 1,2$ the lattice coordinate and vector direction, respectively.  In Eq.~\eqref{eq:lat-ym} $\beta = \tilde{\beta}/a^2$ is the dimensionless coupling.  For the gauge-matter coupling term we have,
\begin{equation}
    S_{\Phi} = -\frac{\kappa}{2} \sum_{x} \sum_{\mu = 1}^{2} \Phi_{x+\hat{\mu}}^{\dagger} U_{x, \mu} \Phi_{x}.
\end{equation}
The $\Phi$ field can be re-expressed in terms of a $2 \times 2$ matrix \cite{montvay_munster_1994} and the gauge-matter term becomes,
\begin{equation}
    S_{\Phi} = -\frac{\kappa}{2} \sum_{x} \sum_{\mu = 1}^{2} \Tr \left[\phi_{x+\hat{\mu}}^{\dagger} U_{x, \mu} \phi_{x}\right],
\end{equation}
where $\phi$ is now a $2 \times 2$ Hermitian matrix.  Since $\phi_{x}^{\dagger} \phi_{x} = \rho_{x}^2 \mathbbm{1}$, $\phi_{x}$ can be written as $\phi_{x} = \rho_{x} \alpha_{x}$ with $\rho_{x} \in \mathbb{R}$, $\rho_{x} \geq 0$, and $\alpha_{x} \in SU(2)$.  This expresses $\phi_{x}$ in terms of the Higgs ($\rho_{x}$) and Goldstone ($\alpha_{x}$) modes, respectively. This allows the gauge-matter term to be again re-written as,
\begin{equation}
\label{eq:gauge-matter}
    S_{\Phi} = -\frac{\kappa}{2} \sum_{x} \sum_{\mu = 1}^{2} \rho_{x+\hat{\mu}} \rho_{x} \Tr \left[ \alpha_{x+\hat{\mu}}^{\dagger} U_{x, \mu} \alpha_{x}\right].
\end{equation}
Because the potential for $\Phi_{x}$ only couples same-site fields it only involves the Higgs mode,
\begin{equation}
\label{eq:potential}
    V = \sum_{x} \rho_{x}^{2} + \lambda(\rho_{x}^{2} - 1)^2.
\end{equation}
The partition function for this model is then,
\begin{equation}
    Z = \int D[U] D[\rho] D[\alpha] e^{-S_{g}-S_{\Phi}-V}
\end{equation}
where the integration over $U$ and $\alpha$ is the $SU(2)$ Haar measure, and the integration measure over $\rho$ is given by $\rho_{x}^{3} d\rho_{x}$ over [0,$\infty$).

\section{Tensor Construction}
\label{sec:tensor-con} 
Starting with the lattice action composed from Eqs.~\eqref{eq:lat-ym},~\eqref{eq:gauge-matter}, and~\eqref{eq:potential},
\begin{align}
\label{initaction}
\nonumber
    S[V, \rho] &= S_{g}[U_{pl}] + 
    \sum_{x} \left \{ \rho_{x}^{2} + \lambda (\rho_{x}^{2} - 1)^{2} \right. \\
    &- \left. \frac{\kappa}{2} \sum_{\mu=1}^{2} \rho_{x+\mu} \rho_{x} \Tr \left[\alpha^{\dag}_{x+\mu}U_{x \mu} \alpha_{x}\right] \right\}
\end{align}
we work in the limit $\lambda \rightarrow \infty$, which forces $\rho \rightarrow 1$ leaving only the Goldstone modes.  The action can be simplified and the Goldstone modes removed by making a gauge transformation.  We choose the transformation
\begin{equation}
    U_{x \mu} \rightarrow U'_{x \mu} = \alpha_{x+\mu}^{\dagger} U_{x\mu} \alpha_{x}
\end{equation}
which only changes the $\kappa$ term.  It now has the form
\begin{equation}
    \frac{\kappa}{2} \sum_{\mu=1}^{2} \Tr [\alpha^{\dag}_{x+\mu}U_{x \mu} \alpha_{x}] \rightarrow \frac{\kappa}{2} \sum_{\mu=1}^{2} \Tr [U_{x \mu}]
\end{equation}
where the prime has been removed for convenience.

Each term in the action is a class function and we can expand the partition function in terms of characters of the gauge group. In general
\begin{equation}
    f(X \Tr[V]) = \sum_{r=0}^{\infty} F_{r}(X) \chi^{r}(V)
\end{equation}
where $F_{r}$ are coefficients of the orthogonal characters of the group, $\chi^r$, and $r$ labels the irreducible representations of the group. 
This is the analog of a Fourier series representation when the group is $O(2)$ or $U(1)$.  The reason this is useful
is that the trace of the matrix representation of the product of group elements, is equal to the trace of a product of matrix representations of group elements, \emph{i.e.}
\begin{align}
    \chi^{r}(U_1 U_2 U_3 \ldots U_n) &= D^{r}_{n n}(U_1 U_2 U_3 \ldots U_n) \nonumber \\
    &= D^{r}_{a b}(U_1) D^{r}_{bc}(U_2) \ldots D^{r}_{z a}(U_n)
\end{align}
with $D^{r}_{mn}(g)$ the matrix representation of the group element $U$ in the irreducible representation $r$, and a sum over repeated indices.

Then the next step is to expand
the Boltzmann weights in terms of characters \cite{DROUFFE19831},
\begin{align}
    e^{-S_{g}} &= \exp \left[ \frac{\beta}{2} \sum_{x} \Tr[U_{x,1} U_{x+\hat{1},2} U^{\dag}_{x+\hat{2},1} U^{\dag}_{x,2}] \right] \nonumber  \\
    &= \prod_{x} \sum_{r} F_{r}(\beta) \chi^{r}(U_{x,1} U_{x+\hat{1},2} U^{\dag}_{x+\hat{2},1} U^{\dag}_{x,2})
\end{align}
and
\begin{align}
    e^{-S_{\Phi}} &= \exp\left[ \frac{\kappa}{2} \sum_{x}\sum_{\mu=1}^{2}
    \Tr [U_{x, \mu}] \right] \nonumber \\
    &=
    \prod_{x, \mu} \sum_{r} F_{r}(\kappa)
    \chi^{r}(U_{x, \mu}).
\end{align}
with
\begin{equation}
    F_{r}(z) = 2 d_{r} \frac{I_{2r+1}(z)}{z}, \quad d_{r} = 2r+1.
\end{equation}
Now, the character of the product can be broken up into
the trace of the product of the matrix representations of the individual elements,
\begin{align}
\label{eq:plaq_trace}
\nonumber
    &\chi^{r}(U_{x,1} U_{x+\hat{1},2} U^{\dag}_{x+\hat{2},1} U^{\dag}_{x,2}) \nonumber \\
    &= D^{r}_{a b}(U_{x,1}) D^{r}_{b c}(U_{x+\hat{1},2}) D^{r\dag}_{c d}(U_{x+\hat{2},1}) D^{r\dag}_{d a}(U_{x,2}) \nonumber \\
    &= D^{r}_{a b}(U_{x,1}) D^{r}_{b c}(U_{x+\hat{1},2}) D^{r\ast}_{d c}(U_{x+\hat{2},1}) D^{r\ast}_{a d}(U_{x,2}) 
\end{align}
where summation is meant for repeated indices here, and
\begin{equation}
    \chi^{r}(U_{x, \mu}) = \sum_{n = -\sigma}^{\sigma} D^{r}_{n n}(U_{x, \mu}).
\end{equation}
With the partition function now written in terms of matrices which are located on the links of the lattice and are completely factorized, we can gather all of the link variables that are associated with the same
link and perform the Haar integration over all of the original group element variables. In two dimensions, there are two plaquettes associated with a single link, as well as the additional link variable coming from the gauge-matter coupling term.  Thus there are a total of three
matrices associated with each link on the lattice, and the integral over each link has the form \cite{quantum-angular}
\begin{align}
\label{eq:3dintegral}
    &\sum_{n = -\sigma}^{\sigma} \int dU D^{r_1}_{m_1 n_1}(U) D^{r_2 \dag}_{m_2 n_2}(U) D^{\sigma}_{n n}(U) \nonumber \\
    &= \sum_{n = -\sigma}^{\sigma}
    \frac{1}{d_{r_2}} C^{r_{2} n_2}_{r_1 m_1 \sigma n} C^{r_2 m_2}_{r_1 n_1 \sigma n}.
\end{align}
Here $C^{j m}_{j_1 m_1 j_2 m_2}$ are Clebsch-Gordan coefficients.  For complete clarity we work this
out for both the horizontal (spatial) case and the vertical (temporal) case.  We use $a,b,l,r$ for `above', `below', `left', and `right' respectively to denote the index location
with respect to the link.  For a horizontal (spatial) link we find,
\begin{align}
    &\sum_{n = -\sigma}^{\sigma} \int dU D^{r_a}_{m_{al} m_{ar}}(U) D^{r_b \dag}_{m_{br} m_{bl}}(U) D^{\sigma}_{n n}(U) \nonumber \\
    &= \sum_{n = -\sigma}^{\sigma}
    \frac{1}{d_{r_b}} C^{r_b m_{bl}}_{r_a m_{al} \sigma n} C^{r_b m_{br}}_{r_a m_{ar} \sigma n},
\end{align}
while for a vertical (temporal) link we find,
\begin{align}
    &\sum_{n = -\sigma}^{\sigma} \int dU D^{r_l}_{m_{lb} m_{la}}(U) D^{r_r \dag}_{m_{ra} m_{rb}}(U) D^{\sigma}_{n n}(U) \nonumber \\
    & = \sum_{n = -\sigma}^{\sigma}
    \frac{1}{d_{r_r}} C^{r_r m_{rb}}_{r_l m_{lb} \sigma n} C^{r_r m_{ra}}_{r_l m_{la} \sigma n}.
\end{align}

\begin{figure}[t]
    \includegraphics[width=8.6cm]{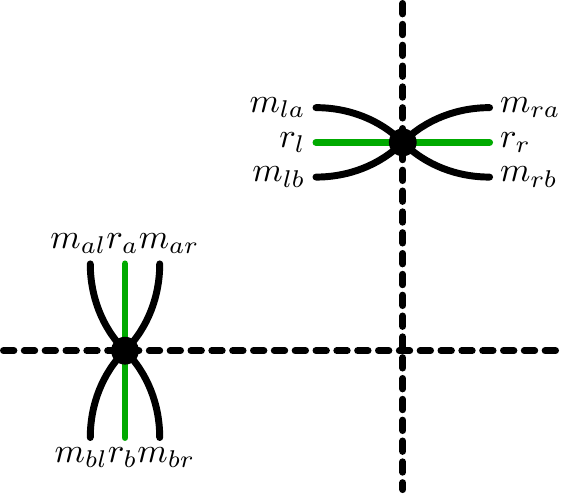}
    \caption{An illustration of two $A$ tensors.  Here the dashed lines represent the original links of the lattice.  The $A$ tensors have two indices, each of which is a product state of three indices.  The subscripts `a', `b`, `l', and `r' denote the relative position of the index corresponding to `above', `below', `left' and 'right'.}
\label{fig:Atensors}
\end{figure}

Instead of a simple flux rule connecting the link and its two neighboring plaquettes, we have the Clebsch-Gordan coefficients connecting
the three representations, which are weighted by $F_{r}(\kappa )$.
A tensor at the link would then have the form
\begin{align}
    &A^{(s)}_{(r_a m_{al} m_{ar})(r_b m_{bl} m_{br})}(\kappa) = \nonumber \\
    &\frac{1}{d_{r_b}} 
    \sum_{\sigma = 0}^{\infty} \sum_{n = -\sigma}^{\sigma} F_{\sigma}(\kappa )C^{r_b m_{bl}}_{r_a m_{al} \sigma n} C^{r_b m_{br}}_{r_a m_{ar} \sigma n}.
\end{align}
on the spatial links, and
\begin{align}
    &A^{(\tau)}_{(r_l m_{la} m_{lb})(r_r m_{ra} m_{rb})}(\kappa) = \nonumber \\
    &\frac{1}{d_{r_r}} 
    \sum_{\sigma = 0}^{\infty} \sum_{n = -\sigma}^{\sigma} F_{\sigma}(\kappa )C^{r_r m_{rb}}_{r_l m_{lb} \sigma n} C^{r_r m_{ra}}_{r_l m_{la} \sigma n}.
\end{align}
on the temporal links.  Here the notation $(r \, m \, m')$ is defined as the product state $r\otimes m \otimes m'$. The Clebsch-Gordan coefficients enforce that $m_{rb} - m_{lb} = n = m_{ra} - m_{la}$ which enables us to sum over $n$.
In addition, by the triangle
inequalities on $r_1$, $r_2$, and $\sigma$, we can rewrite this as
\begin{align}
\label{eq:atenh}
\nonumber
    &A^{(s)}_{(r_a m_{al} m_{ar})(r_b m_{bl} m_{br})}(\kappa) = \\ \nonumber
    &\frac{1}{d_{r_b}} 
    \sum_{\sigma = |r_b-r_a|}^{r_b+r_a} F_{\sigma}(\kappa )C^{r_b m_{bl}}_{r_a m_{al} \sigma (m_{bl}-m_{al})} \times \\
    &C^{r_b m_{br}}_{r_a m_{ar} \sigma (m_{bl}-m_{al})}.
\end{align}
and
\begin{align}
\label{eq:atenv}
\nonumber
    &A^{(\tau)}_{(r_l m_{la} m_{lb})(r_r m_{ra} m_{rb})}(\kappa) = \\ \nonumber
    &\frac{1}{d_{r_r}} 
    \sum_{\sigma = |r_r-r_l|}^{r_r+r_l} F_{\sigma}(\kappa )C^{r_r m_{rb}}_{r_l m_{lb} \sigma (m_{rb} - m_{lb})} \times \\
    &C^{r_r m_{ra}}_{r_l m_{la} \sigma (m_{rb} - m_{lb})}.
\end{align}
In this construction, the conventions we are using based on the Wigner $D$-matrices is the following: For $C^{j m}_{j_1 m_1 j_2 m_2}$, left and above correspond to subscript 1, and right and below correspond to no subscript. These conventions match the Abelian case \cite{abelian-higgs}, and give a charge
$m_{\text{right}}-m_{\text{left}}$ and $m_{\text{below}}-m_{\text{above}}$. A visual representation of these tensors can be seen in Fig. \ref{fig:Atensors}.

\begin{figure}[t]
    \includegraphics[width=8.6cm]{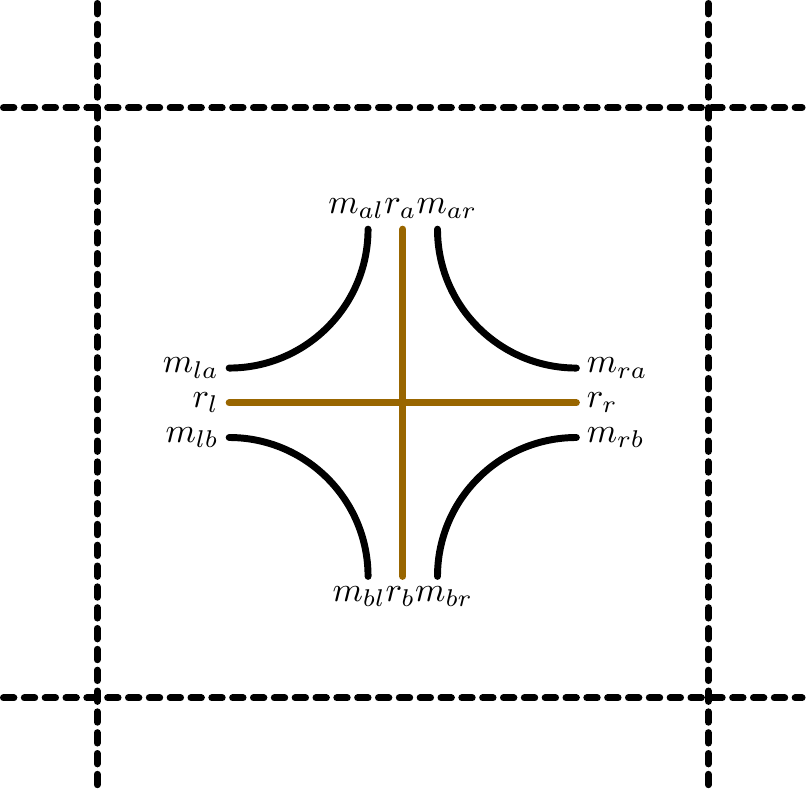}
    \caption{An illustration of a $B$ tensor. The black lines on the outside are Kronecker deltas, and the brown cross is a diagonal tensor which forces all four $r$s to be identical.  The dashed lines are the original links of the lattice.  The subscripts `a', `b`, `l', and `r' denote the relative position of the index corresponding to `above', `below', `left' and 'right'.}
\label{fig:Btensor}
\end{figure}

Similar to the Abelian case \cite{abelian-higgs,abelian-ploop}, at each plaquette we have a factor of
$F_{r}(\beta)$ which demands that the incoming representations all are the same.  However,
in addition the magnetic quantum numbers (the matrix indices) must be closed around the plaquette as they were in the original formulation.  Then, not only is there a demand that all four incoming
representations are the same, but that neighboring magnetic quantum numbers are the same.
On the plaquette we have the tensor,
\begin{widetext}
\begin{equation}
\label{eq:btensor}
    B_{(r_l m_{la} m_{lb})(r_r m_{ra} m_{rb})(r_a m_{al} m_{ar})(r_b m_{bl} m_{br})} =
    \begin{cases}
        F_{r}(\beta) \delta_{m_{la}, m_{al}} \delta_{m_{ar}, m_{ra}} \delta_{m_{rb}, m_{br}} \delta_{m_{bl}, m_{lb}} & \text{if } r_l = r_r = r_a = r_b = r \\
        0 & \text{else.}
    \end{cases}
\end{equation}
\end{widetext}
This tensor forces the surrounding tensors to share a common irreducible representation at that plaquette, and 
directs the magnetic quantum numbers around the plaquette loop.  This can be seen in Fig. \ref{fig:Btensor}, where the Kronecker deltas enforce the same trace as in Eq.~\eqref{eq:plaq_trace}.

\begin{figure}
    \centering
    \includegraphics[width=8.6cm]{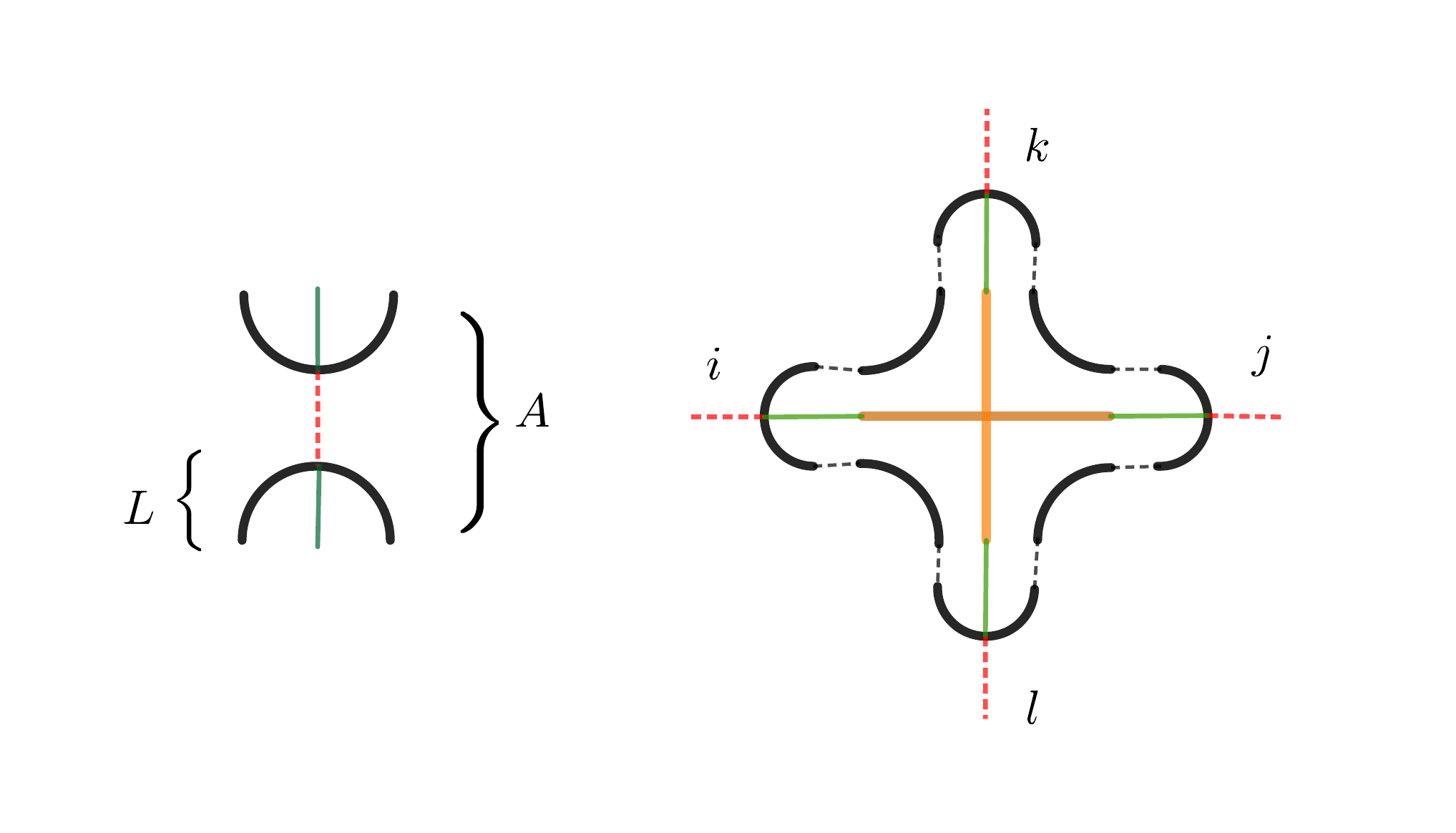}
    \caption{\textbf{Left:} A graphical interpretation of the matrix factorization of the $A$ tensor into the product of two identical matrices, $A = L L^{T}$.  The dashed line in between represents an intermediate sum over states. \textbf{Right:} The fundamental tensor $T_{ijkl}$ is built by contracting four $L$ matrices together with a single $B$ tensor at a plaquette.  Forming this tensor sums over all the $r$ and $m$ variables on the lattice, and instead one is left with the sums over the intermediate indices from the $L$ matrices.}
    \label{fig:A-LL-T}
\end{figure}

The final partition function is now written as
\begin{equation}
\label{eq:abtensor-pf}
    Z(\beta, \kappa) = \Tr\left[ \left( \prod_{x} B^{(x)}(\beta) \right) \left( \prod_{x, \mu} A^{(x,\mu)}(\kappa) \right) \right]
\end{equation}
where the product over $B$ tensors is over each plaquette associated with a site, and the product over $A$ tensors is over every link.  The trace is over all tensor indices, which have been suppressed.  This can be further simplified by defining a single local tensor built from the $A$ and $B$ tensors.  Since the $A$ tensor has two product-state indices, it can be interpreted as a matrix.  This matrix allows for a factorization of the form $A = L L^{T}$.  Since each plaquette is bounded by four links, and each link is associated with a single $A$ tensor, we can assign four $L$ matrices to a single $B$ tensor throughout the lattice, and define a fundamental tensor $T$,
\begin{equation}
    T_{ijkl}(\beta, \kappa) = \sum_{\alpha,\beta,\gamma,\delta} B_{\alpha \beta \gamma \delta}(\beta) L_{\alpha i} L_{\beta j} L_{\gamma k} L_{\delta l}(\kappa).
\end{equation}
A graphical representation of this fundamental tensor, along with the decomposition of the $A$ tensor, can be seen in Fig.~\ref{fig:A-LL-T}.
By contracting the $T$ tensor with itself repeatedly in the shape of a square lattice, we reproduce the partition function from Eq.~\eqref{eq:abtensor-pf}.  The partition function can now be written as,
\begin{equation}
    Z(\beta, \kappa) = \Tr\left[ \prod_{x} T^{(x)}(\beta, \kappa) \right]
\end{equation}
where again the trace is over all tensor indices.

 From this point on we use normalized coefficients, unless otherwise stated, of the form $f_{r}(z) = F_{r}(z)/F_{0}(z)$.  
 For this model, these functions have the following large, and small argument behavior,
\begin{align}
    f_r(z) &= d_r \left(1 - \frac{2r(2r + 2)}{2z} \right) + \mathcal{O}(z^{-2}) \text{ as } z \rightarrow \infty \nonumber \\
    &= d_r \left( 1 - \frac{\lambda_r(\lambda_r + 2)}{2z} \right) + \mathcal{O}(z^{-2})
\end{align}
with $\lambda_r = 2r$, and
\begin{equation}
    f_r(z) = z^{\lambda_r} \frac{4^{-r} d_r}{\Gamma(\lambda_r +2)} + \mathcal{O}(z^{\lambda_r+2}) \text{ as } z \rightarrow 0.
\end{equation}
Looking at the large argument behavior we see that the leading order is just the eigenvalue for the quadratic Casimir operator for $SU(2)$.

\section{The continuum limit, $\beta = 0$, and $\kappa = 0$}
\label{sec:limits}

In this section we discuss how the continuum limit is approached from the lattice model, as well as some limiting cases for the model.  By looking at the continuum action for the pure Yang-Mills case, Eq. \eqref{eq:ym-action}, we see that $\tilde{\beta}$ must have dimensions of length squared.  This indicates that the important ratio is $\tilde{\beta}/V_{\text{phys.}}$
Thus the continuum, fixed physical volume limit is set when $\tilde{\beta}/V_{\text{phys.}} = c$, with $c$ some constant, 
and $V_{\text{phys.}}$ a dimensionful spacetime volume.
This limit can be calculated on the lattice if we take $\beta/N_{s}N_{\tau} = c$ with $\beta \rightarrow \infty$ and $N_{s}, N_{\tau} \rightarrow \infty$. In this limit, the gauge coupling, $g$, becomes arbitrarily weak as the lattice volume is taken infinitely large.

\subsection{$\beta = 0$ limit}

In this limit the partition function becomes a product of one-link integrals,
\begin{align}
\label{eq:beta0}
\nonumber
    Z &= \int \mathcal{D}[U] e^{\frac{\kappa}{2} \sum_{x,\mu} \chi^{\frac{1}{2}}(U_{x,\mu})} \\ \nonumber
    &= \prod_{x,\mu} \int dU_{x,\mu} e^{\frac{\kappa}{2}\chi^{\frac{1}{2}}(U_{x,\mu})} \\ \nonumber
    &= \prod_{x,\mu} \frac{1}{\pi} \int_{0}^{2\pi} d\theta_{x,\mu} \sin^{2}\left(\frac{\theta_{x,\mu}}{2}\right) \exp\left[ \frac{\kappa \sin\theta_{x,\mu}}{2\sin(\theta_{x,\mu}/2)} \right] \\
    &= \prod_{x,\mu} \left( 2 \frac{I_{1}(\kappa)}{\kappa} \right) = \left( \frac{2 I_{1}(\kappa)}{\kappa} \right)^{2 N_{s} N_{\tau}} \nonumber \\
    & = \left( F_{0}(\kappa) \right)^{2 N_{s} N_{\tau}}.
\end{align}
By setting $\beta = 0$ in the tensor formulation, the $B$ tensors immediately imply that the only non-zero contributions to the partition function will be the $r=0$ representation on each plaquette.  This enforces that the link tensors also simplify to the $\sigma=0$ representation.  This immediately gives Eq.~\eqref{eq:beta0}.


\subsection{$\kappa = 0$ limit}
\label{sec:kappa0}

In this limit the partition function becomes that of pure $SU(2)$ gauge theory.  This model is also solvable and the simplest way to see this is using the character expansion from before.  Since the matter-gauge coupling is turned off, there are only two link variables to integrate per link.  The integral associated with each link has the form,
\begin{equation}
\label{eq:2dintegral}
    \int dU_{x,\mu} D^{r}_{m n}(U) D^{r' \dagger}_{m' n'}(U) =
    \frac{1}{d_r} \delta_{r r'} \delta_{m n'} \delta_{n m'},
\end{equation}
which is Eq. \eqref{eq:3dintegral} without the additional Wigner $D$-matrix from the matter-gauge term.  This forces each irreducible representation of $SU(2)$ on each plaquette to be the same across the whole lattice, and for each representation there is a degeneracy of $d_r$.  The partition function takes the form,
\begin{equation}
    Z = (F_0(\beta))^{N_{s} N_{\tau}} \sum_{r} \left( \frac{f_{r}(\beta)}{d_r} \right)^{N_{s} N_{\tau}}.
\end{equation}
In the tensor language when $\kappa = 0$, the only contribution from the $A$ tensors is the $\sigma=0$ representation.  This forces the Clebsch-Gordan coefficients to simplify into Kronecker deltas, reproducing the integral from Eq. \eqref{eq:2dintegral}.

It's useful to consider the physical interpretation of the half-integer $r$ numbers.  Notice that the $r$ fields are associated with the plaquettes of the original lattice and seem to naturally play the role of dual variables.  In fact, looking at the $A$ tensors from Eqs.~\eqref{eq:atenh}, and~\eqref{eq:atenv}, the $\sigma$ field is summed-out of the partition function, leaving only a model in terms of the plaquette variables, and the $m$ variables.

To further understand the $r$ fields, consider the limit of $\beta \rightarrow \infty$, which is similar to the continuum limit.  In this limit the partition function becomes,
\begin{equation}
    Z \approx (F_{0}(\beta))^{N_s N_\tau} \sum_r \left( 1 - \frac{\lambda_r (\lambda_r +2)}{2\beta} \right)^{N_s N_\tau}.
\end{equation}
If we further consider that only the leading order is important, we can rewrite the partition function as,
\begin{equation}
    Z \approx (F_{0}(\beta))^{N_s N_\tau} \sum_r \exp\left[-a\frac{U}{2}\sum_{i=1}^{N_s} \lambda_{r,i} (\lambda_{r,i} +2) \right]^{N_\tau},
\end{equation}
with $U \equiv 1/a\beta$ where $a$ is the lattice spacing.  We see this this nothing more than the Hamiltonian for the pure Yang-Mills case.  The Hamiltonian in this case is
\begin{equation}
\label{eq:ym-ham}
    H = \frac{U}{2}\sum_{i=1}^{N_s} \mathcal{C}^2_i = \frac{U}{2}\sum_{i=1}^{N_s} \vec{E}_i \cdot \vec{E}_i
\end{equation}
with $\mathcal{C}^2$ the quadratic Casimir operator for $SU(2)$ and $\vec{E}$ the electric field.  Therefore we see that the $r$ fields on the lattice are the discrete quantum numbers of the electric field of the non-Abelian gauge field.

Note that since there is only one direction to travel in space, the electric field is the same everywhere, and so the quadratic Casimir element only takes on one value across all of spacetime.  The energy density then takes simple values,
\begin{align}
\label{eq:casimir-eigs}
\nonumber
    \frac{E_{r}}{N_{s}} &= \frac{U}{2}\bra{r}\mathcal{C}^{2}\ket{r} \\
    \nonumber
    &= \frac{U}{2}\lambda_r (\lambda_r + 2) \\
    &= \frac{U}{2}(0, 3, 8, \ldots ) \text{ for } r=0,\frac{1}{2},1,\ldots.
\end{align}

\subsection{The mass gap}

\begin{figure}[t]
    \centering
    \includegraphics[width=8.6cm]{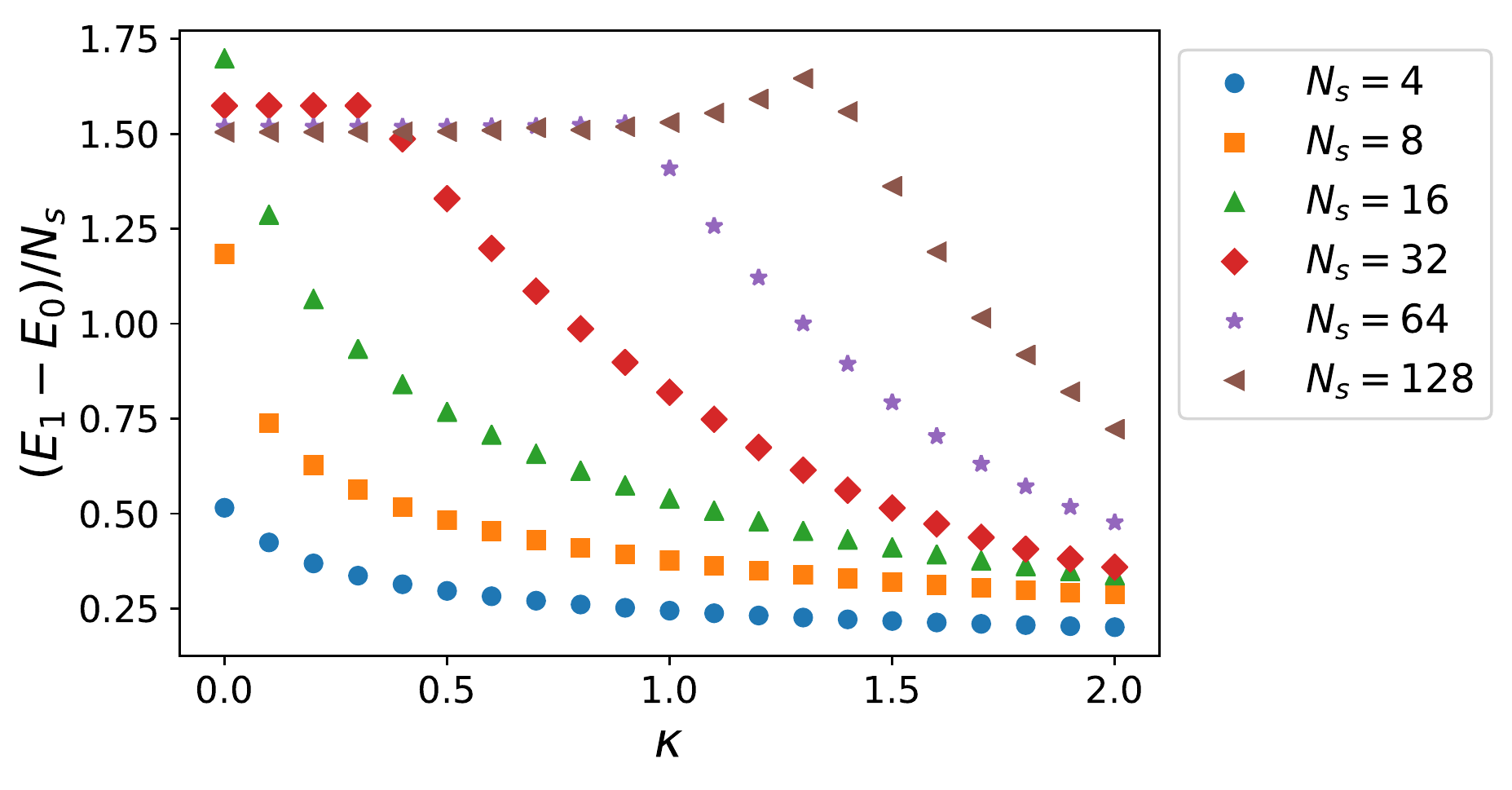}
    \caption{The mass gap density as a function of $\kappa$ while taking the continuum limit.  Here $\beta / N_{s}^{2} = 0.01$ is held fixed as the volume is increased.  We see the gap approaches the correct $\kappa = 0$ value of $3/2$ as the continuum limit is taken.}
    \label{fig:egap_0p01}
\end{figure}

Using the tensor construction from Sec.~\ref{sec:tensor-con} one can construct a transfer matrix, $\mathbb{T}$, by contracting tensors only along a time-slice.  Here we use the HOTRG to approximate the tensor contractions \cite{Xie:2012}.  By blocking along a time-slice and constructing an approximate transfer matrix, we can diagonalize this matrix and extract the relative eigenvalues of the Hamiltonian through the relation,
\begin{equation}
    \mathbb{T} = e^{-a H},
\end{equation}
where $a$ is the temporal lattice spacing, and $H$ is the Hamiltonian.  This allows us to calculate the mass gap, $M$, in units of $U$ in the continuum from,
\begin{equation}
    \frac{M}{U} = \beta \ln\left( \frac{\lambda_1}{\lambda_{0}}  \right) = 
    \frac{E_{1} - E_{0}}{U}
\end{equation}
where $\lambda_{n}$ are the eigenvalues of the transfer matrix ordered from largest to smallest as $\lambda_0$, $\lambda_1$, \ldots  This can be seen for the mass gap as a function of $\kappa$ in Fig.~\ref{fig:egap_0p01}.  Looking at the $\kappa \rightarrow 0$ limit in Fig.~\ref{fig:egap_0p01}, in units of $U$, this matches Eq.~\eqref{eq:casimir-eigs} for the first excited state.
This figure shows the progression to the continuum limit as the size of the lattice increases, keeping the ratio $\beta / N_{s}^{2}$ fixed to a constant, 0.01.

This is a pleasant property of the TRG.  One is able to calculate approximate, relative eigenvalues of the transfer matrix, and hence of the Hamiltonian for the system.  Typically the eigenvalue spectrum must be calculated in other ways in sampling methods, \emph{e.g.} Monte Carlo, while in the TRG one has direct access to their relative size.

\section{Observables and comparison with Monte Carlo}
\label{sec:mc-compare}

To check the tensor formulation, as well as explore the model in greater detail, we computed expectation values of operators using the TRG and Monte Carlo simulations and compared them. The MC simulations implemented the Hybrid Monte Carlo algorithm maintaining about 70-80 \% acceptance for all the ensembles. The MC runs were carried out for at least 50000 molecular dynamics time units (MDTU) while measuring the expectation values every two MDTU. To compute the averages, we set a thermalization cut of at least 10000 MDTU ($\emph{i.e.}$ 5000 measurements) for all ensembles. The resulting sample average and errors were calculated using the standard jackknife binning procedure.  

The observables we consider are the average plaquette, $\langle p \rangle$, the expectation value of the gauge-matter term, $\langle L_{\phi} \rangle$, and its susceptibility, $\chi_{L_{\phi}}$, and finally the Polyakov loop, $\langle P \rangle$ and its correlation function $G_{P P^{\dagger}}$.  These expressions are given by
\begin{align}
\label{eq:avg-plaq}
    \langle p \rangle &= \frac{1}{N_{s} N_{\tau}} \frac{\partial \ln Z}{\partial \beta}
\end{align}
and
\begin{equation}
\label{eq:lphi}
    \langle L_{\phi} \rangle = \frac{1}{N_{s}N_{\tau}} \frac{\partial \ln Z}{\partial \kappa},
\end{equation}
with $L_{\phi} = \sum_{x,\mu} \Tr[U_{x,\mu}]/2$, and
\begin{equation}
\label{eq:gl-sup}
    \chi_{L_{\phi}} = \frac{1}{N_{s} N_{\tau}} \frac{\partial^{2} \ln Z}{\partial \kappa^{2}} = \frac{1}{N_{s} N_{\tau}} \left\langle (L_{\phi} - \langle L_{\phi} \rangle)^{2} \right\rangle.
\end{equation}

\begin{figure}
    \centering
    \includegraphics[width=8.6cm]{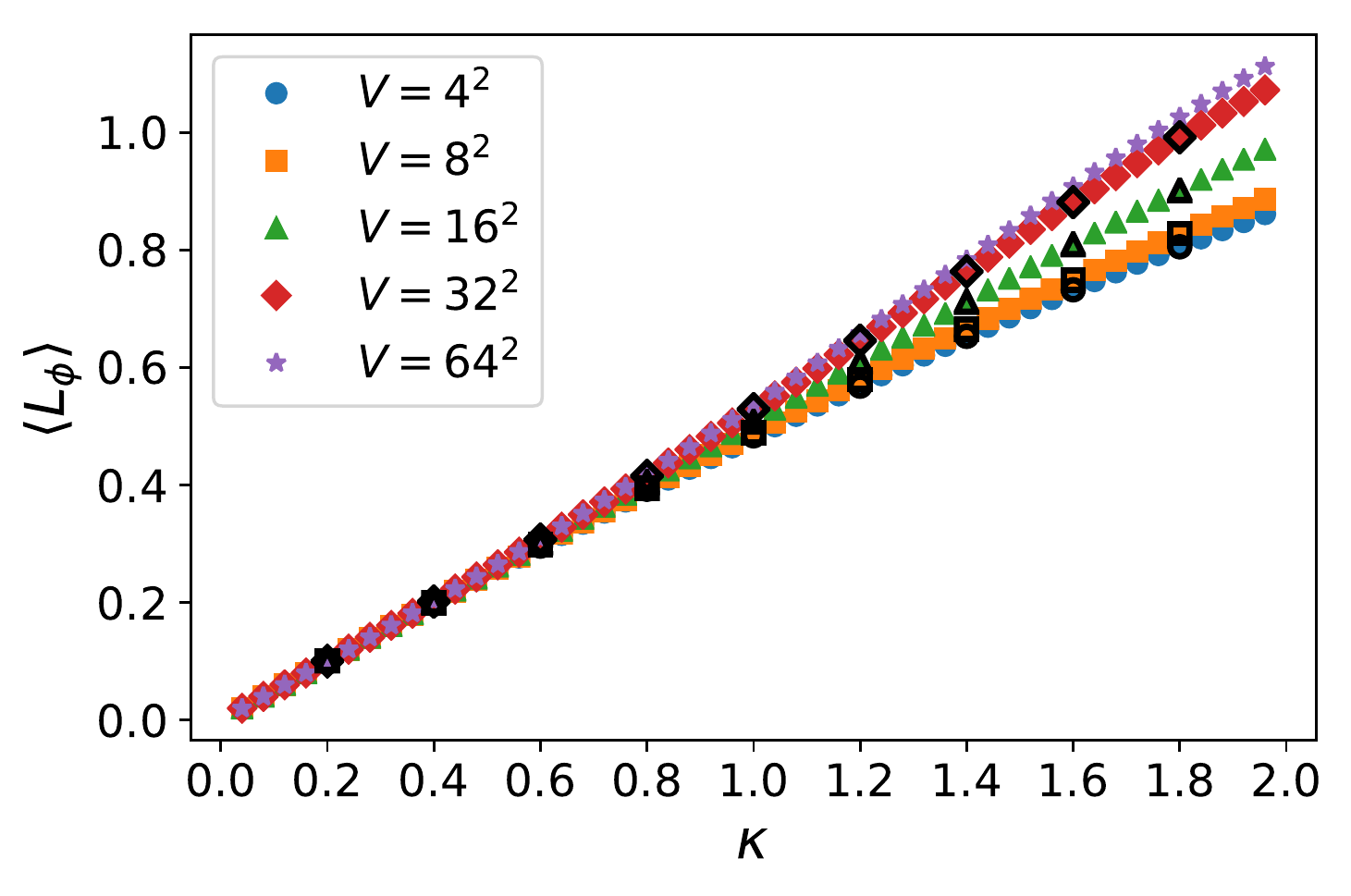}
    \caption{Calculation of the average trace of a link variable, Eq.~\eqref{eq:lphi}, using the TRG and comparing with MC.  Here the continuum limit is approached by keeping the ratio $\beta / N_{s} N_{\tau}$ fixed at $0.01$ while increasing the volume.  The colored markers are from the TRG calculations which were done with a local state space truncation at $r_{\text{max}}=1$ and $D_{\text{bond}}=50$.  The Monte Carlo data are the black hollow markers.}
    \label{fig:lphi}
\end{figure}

In Fig.~\ref{fig:lphi} we have computed $\langle L_{\phi} \rangle$ using both MC and the TRG and compared them while taking the continuum limit.
In this section we have taken $\beta / V = c = 0.01$ throughout, however other $c$ values were tested.  Smaller values of $c$ simply correspond to a slower convergence to the continuum limit.
Here the initial truncation on the TRG state space was at $r_{\text{max}} = 1$, limiting the local Hilbert space to 14 states (one from the trivial representation, four from the fundamental representation, and nine from the adjoint).  The final number of states kept was $D_{\text{bond}} \sim 50$.  This was typical for the runs in this paper.  We see for small $\kappa$ rapid convergence to the continuum limit; however, for larger $\kappa$ there is convergence to the continuum limit for sufficiently large volumes.

\begin{figure}[t]
    \centering
    \includegraphics[width=8.6cm]{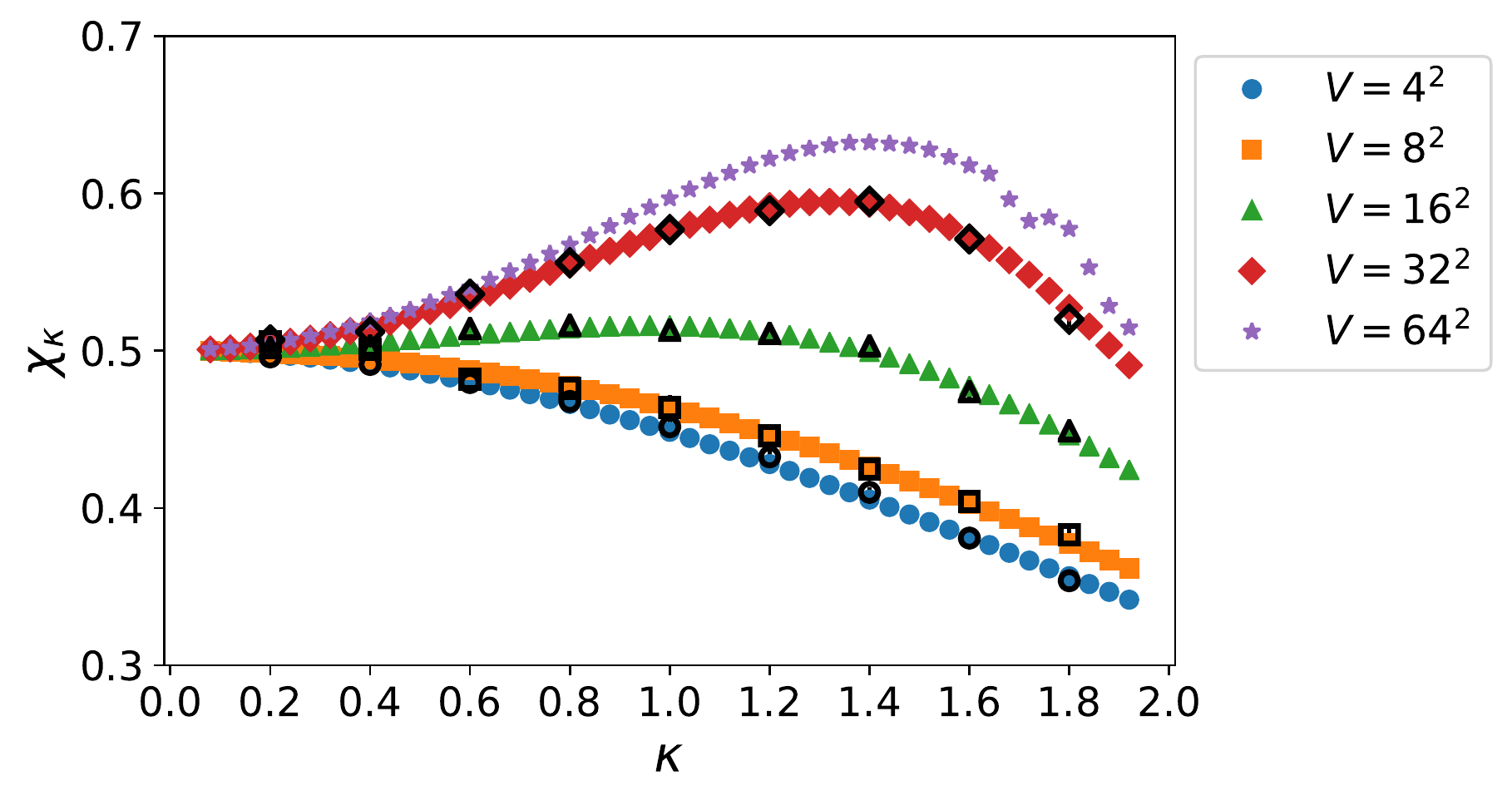}
    \caption{The gauge-link susceptibility, Eq.~\eqref{eq:gl-sup}, as a function of $\kappa$ as one approaches the continuum-limit.  Here $\beta / V = 0.01$ is held fixed as the volume is increased.  We see convergence as the volume gets sufficiently large indicating a crossover behavior.  The colored markers are from the TRG calculations which were done with a local state space truncation at $r_{\text{max}}=1$ and $D_{\text{bond}}=50$. The Monte Carlo data are the black hollow markers.}
    \label{fig:kappa-sup-0p01}
\end{figure}

The susceptibility $\chi_{L_{\phi}}$ can be seen in Fig.~\ref{fig:kappa-sup-0p01} as one takes the continuum limit.
We see the peak does not tend to diverge but rather settles as the volume gets large, although the TRG data is somewhat noisy for $\kappa \gtrsim 1.4$ for the larger volumes.  This is indicative of a cross-over behavior which matches previous expectations \cite{Gongyo:2014jfa}.  The small $\kappa$ regime to the left of the peak is associated with the confining regime, with the pure Yang-Mills model in the limit of $\kappa \rightarrow 0$ having confinement.  At larger $\kappa$ to the right of the peak we are in the Higgs regime.

\begin{figure}[t]
    \centering
    \includegraphics[width=8.6cm]{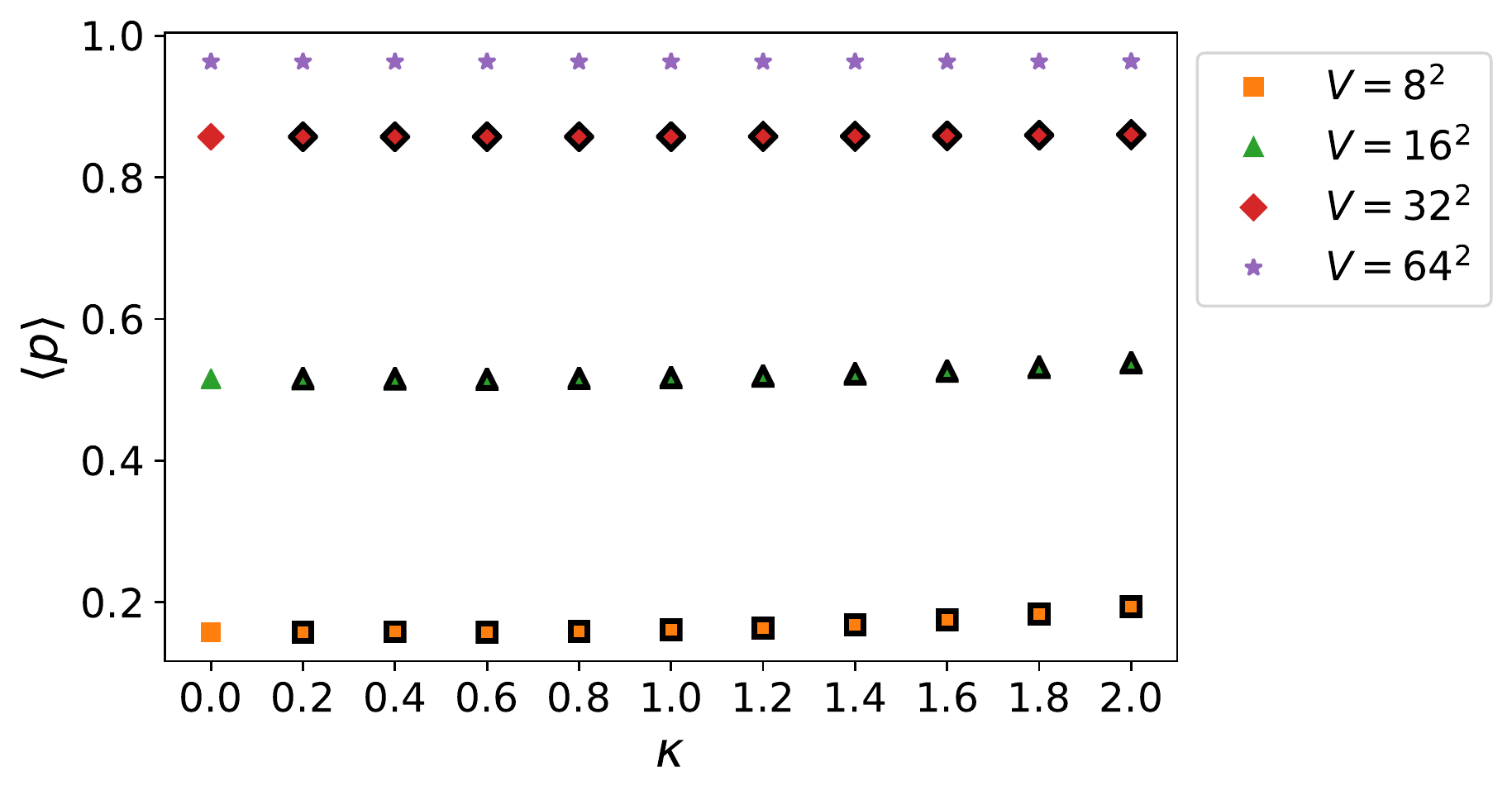}
    \caption{The average plaquette, Eq.~\eqref{eq:avg-plaq}, as a function of $\kappa$ as one approaches the continuum-limit.  Here $\beta / V = 0.01$ is held fixed as the volume is increased.    The colored markers are from the TRG calculations which were done with a local state space truncation at $r_{\text{max}}=1$ and $D_{\text{bond}}=50$.  The Monte Carlo data are the black hollow markers.}
    \label{fig:avg-plaq}
\end{figure}

In Fig.~\ref{fig:avg-plaq} we see the average plaquette as a function of $\kappa$ as one takes the continuum limit.
The average plaquette can be computed directly from $\ln Z$ by taking a numerical derivative as in Eq.~\eqref{eq:avg-plaq}.  As can be seen from the figures, we find good agreement between the two methods.

\subsection*{Polyakov loop and its correlation function}

    The Polyakov loop is defined as,
    \begin{equation}
        P_{x^*} \equiv \Tr \left[ \prod_{n=0}^{N_{\tau}-1} D^{\frac{1}{2}}(U_{x+n\hat{2},2}) \right],
    \end{equation}
    with the gauge fields, $D^{r}$, in the $1/2$, or fundamental, representation and was investigated for the Abelian-Higgs model using a tensor formulation in Refs.~\cite{abelian-ploop}.  When constructing the tensor formulation for the Polaykov loop, this adds an additional group element on a loop of temporal links.  These links have a special integration different from the above.  We use the Clebsch-Gordan series on the first and last $D$-matrices, and then use the series again and integrate for the three remaining $D$-matrices,
    \begin{align}
        &\sum_{n = -\sigma}^{\sigma} \int dU \, D^{r_{l}}_{m_{lb} m_{la}} D^{r_r \dag}_{m_{ra} m_{rb}} D^{\sigma}_{nn}
        D^{\frac{1}{2}}_{ij} = \nonumber \\
        & \sum_{r' m m' n} \frac{1}{d_{r_r}} C^{r_r m_{rb}}_{r' m \sigma n}
        C^{r_r m_{ra}}_{r' m' \sigma n} C^{r' m}_{r_l m_{lb} \frac{1}{2} i}
        C^{r' m'}_{r_l m_{la} \frac{1}{2} j}.
    \end{align}
    If one continues like above and does the $n$ sums at each link we get,
    \begin{align}
        &\sum_{n = -\sigma}^{\sigma} \int dU \, D^{r_{l}}_{m_{lb} m_{la}} D^{r_r \dag}_{m_{ra} m_{rb}} D^{\sigma}_{nn}
        D^{\frac{1}{2}}_{ij} = \nonumber \\
        & \sum_{r' m m'} \frac{1}{d_{r_r}} C^{r_r m_{rb}}_{r' m \sigma (m_{rb} - m)}
        C^{r_r m_{ra}}_{r' m' \sigma (m_{rb} - m)} \times \nonumber \\
        &C^{r' m}_{r_l m_{lb} \frac{1}{2} i}
        C^{r' m'}_{r_l m_{la} \frac{1}{2} j},
    \end{align}
    and then doing the $m$ and $m'$ sums we get
    \begin{align}
        &\sum_{n = -\sigma}^{\sigma} \int dU \, D^{r_{l}}_{m_{lb} m_{la}} D^{r_r \dag}_{m_{ra} m_{rb}} D^{\sigma}_{nn}
        D^{\frac{1}{2}}_{ij} = \nonumber \\
        & \sum_{r' = |\frac{1}{2} - r_l|}^{\frac{1}{2}+r_l}
         \frac{1}{d_{r_r}} C^{r_r m_{rb}}_{r' (m_{lb}+i) \sigma (m_{rb} - m_{lb}-i)} \times \nonumber \\
        &C^{r_r m_{ra}}_{r' (m_{la}+j) \sigma (m_{rb} - m_{lb}-i)} C^{r' (m_{lb}+i)}_{r_l m_{lb} \frac{1}{2} i}
        C^{r' (m_{la}+j)}_{r_l m_{la} \frac{1}{2} j}.
    \end{align}
    With this expression for the integral of four Wigner $D$-matrices, we can write down the ``impure'' $A$ tensor associated with those temporal links which contain the Polyakov loop.
    \begin{align}
    \label{eq:ploopA}
    \nonumber
        &\tilde{A}^{(\tau)}_{(r_l m_{la} m_{lb})(r_r m_{ra} m_{rb}) i j}(\kappa) = \\ \nonumber
    &\frac{1}{d_{r_r}} \sum_{r' = |\frac{1}{2} - r_l|}^{\frac{1}{2}+r_l} \sum_{\sigma = |r_r - r'|}^{r_r +r'} f_{\sigma}(\kappa)
          C^{r_r m_{rb}}_{r' (m_{lb}+i) \sigma (m_{rb} - m_{lb}-i)} \times \\
        &C^{r_r m_{ra}}_{r' (m_{la}+j) \sigma (m_{rb} - m_{lb}-i)} C^{r' (m_{lb}+i)}_{r_l m_{lb} \frac{1}{2} i}
        C^{r' (m_{la}+j)}_{r_l m_{la} \frac{1}{2} j}.
    \end{align}
    The computation of $\tilde{A}$ for $P^{\dagger}$ is done in a similar fashion using the Clebsch-Gordan series.  The $i$ and $j$ indices of this tensor should be contracted with other $\tilde{A}$ tensors along the temporal direction and then traced over with periodic boundary conditions.
    
    For the Polyakov loop correlation function, we have
    \begin{equation}
        G_{P P^{\dagger}}(d) = \langle P_{0} P^{\dagger}_{d} \rangle.
    \end{equation}
    where $d$ is the separation between loops.  This involves the insertion of two Polyakov loops, each winding in different directions representing a static color and anti-color charged pair.  The calculation of this observable only requires the impure $\tilde{A}$ tensor constructed before.
    The correlation function is defined to be related to the static quark potential through,
    \begin{equation}
        G_{P P^{\dagger}}(d) \simeq \exp\left[ -V(d) / T \right] = \exp[ -a V(d) N_{\tau} ].
    \end{equation}
    where $T$ is the physical temperature.  In the confining regime, one expects a linear potential for $V$, while in the Higgs regime one would expect, after some distance, that the pair breaks and only a constant potential is realized.
    
\begin{figure}
        \centering
        \includegraphics[width=8.6cm]{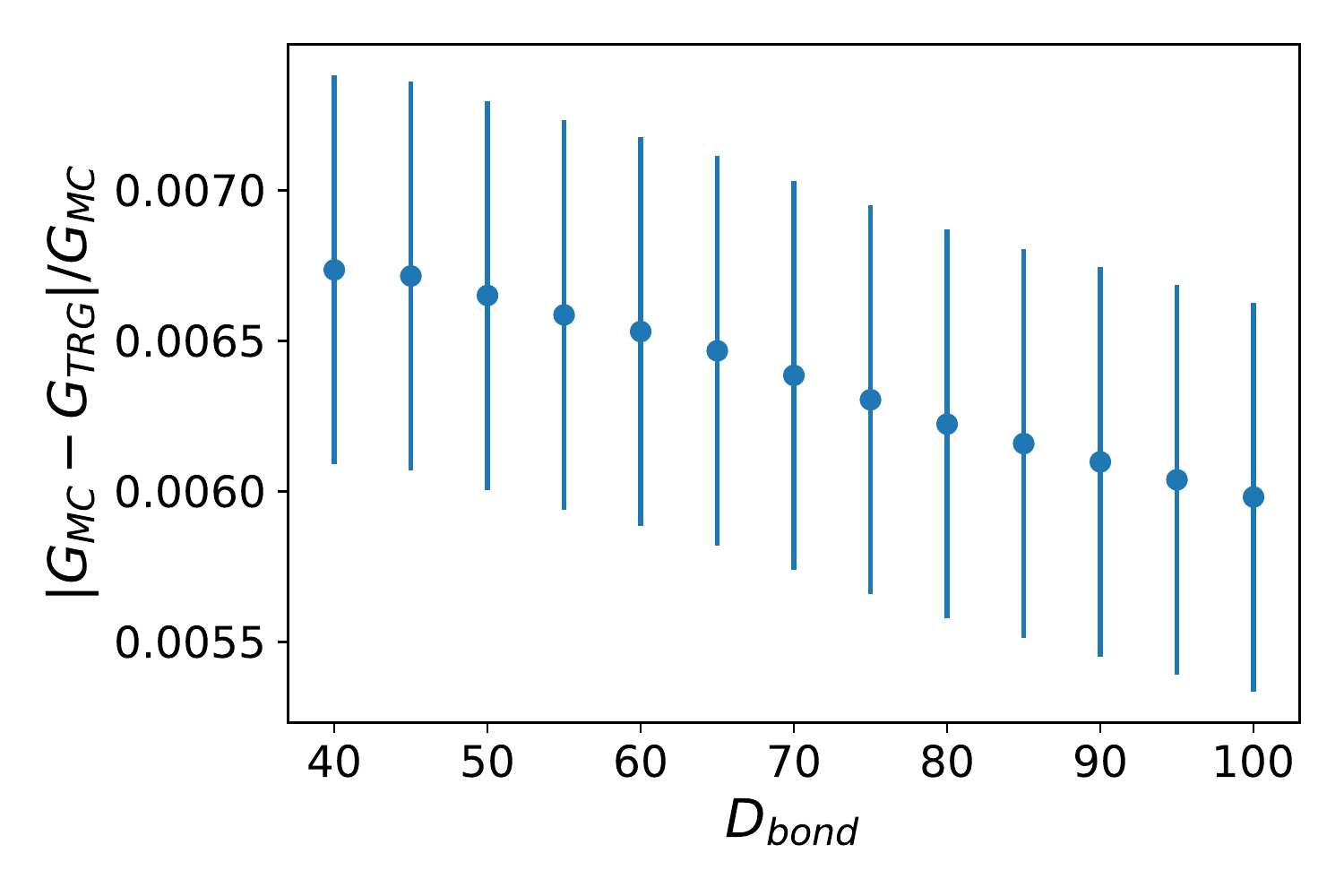}
        \caption{The relative error of the Polyakov loop correlation function between the TRG and MC.  This is for a larger value of $\kappa = 2$, on a $8\times1$ lattice with a separation of $d=4$ between the two loops.  We see the TRG solution converges slowly to the MC value at larger final bond dimension.}
        \label{fig:gmc-trg-compare}
    \end{figure}

    An important computational feature is that the correlation function is suppressed exponentially in the temporal lattice extent.  This makes calculations of this quantity using Monte Carlo methods extremely difficult, since, for even modest lattices, one loses the signal to the noise unless long runs are carried out.  On the other hand, the TRG at its heart is simply a multi-linear algebra calculation, and while state truncation introduces systematic errors, in principle such small numbers are not an issue.  However, comparison of the Polyakov loop and its correlation function in regimes where both the Monte Carlo and the TRG can be used ($N_{\tau} \approx 1$) indicate that at larger $\kappa$ values ($\kappa \gtrsim 1$), the TRG is less quantitatively accurate, but retains correct qualitative features.  In Fig.~\ref{fig:gmc-trg-compare} we show the relative error between a calculation of $G$ using the TRG and MC at a fixed value of $\kappa = 2$, for a separation of $d = 4$ on a $8\times1$ lattice.  We see relatively slow, but consistent convergence to the Monte Carlo number.

\begin{figure}[t]
        \centering
        \includegraphics[width=8.6cm]{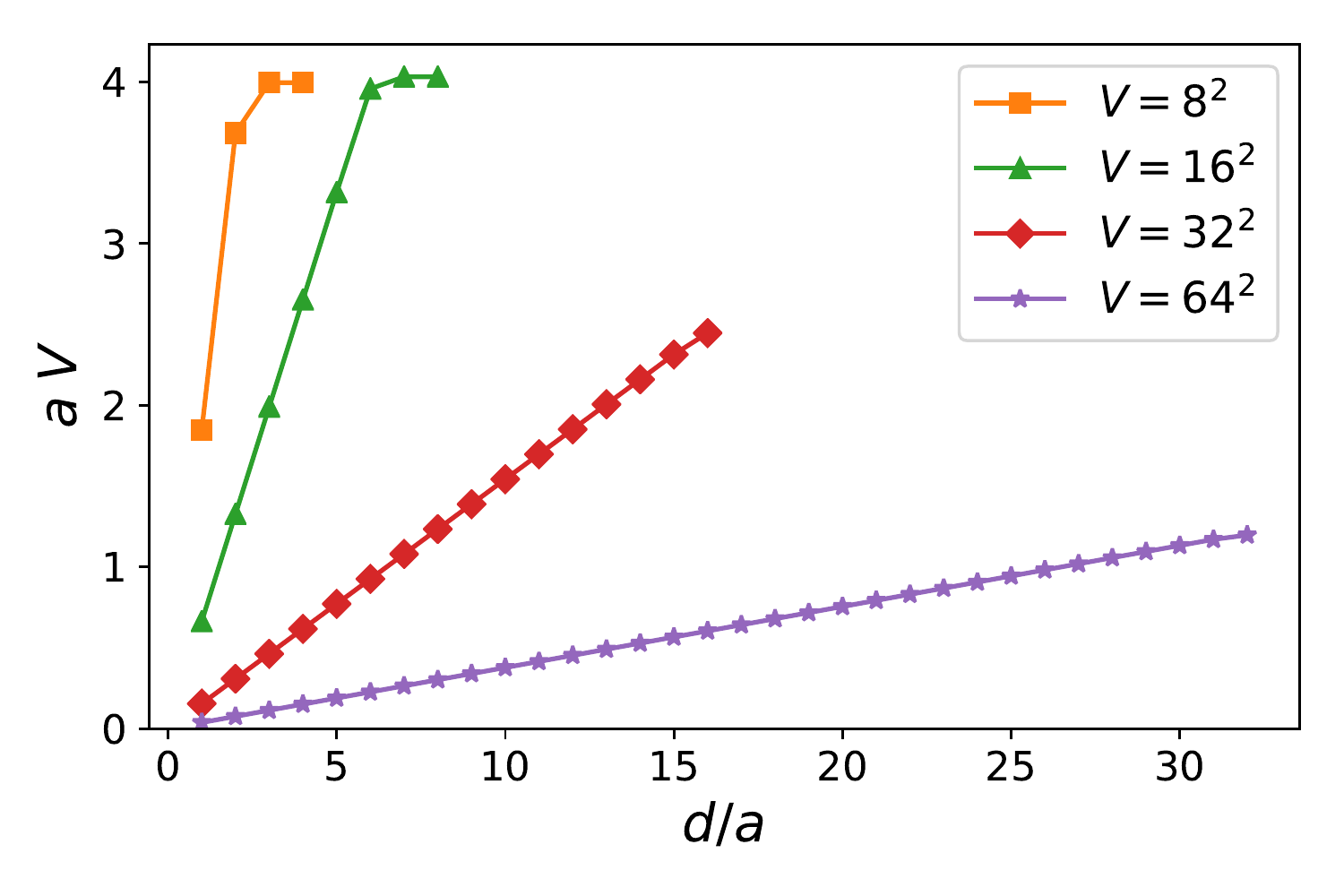}
        \caption{The static charge potential, $V$, from the Polaykov correlation function in the continuum limit.  Here we took $\beta / N_{s}N_{\tau} = 0.01$ fixed as we increased the size of the lattice.  These runs were done with the gauge-matter coupling $\kappa = 1/2$, which is well in the confining regime of the model.  This is characterized by a dominant linear potential across the whole system.}
        \label{fig:pcorr-k0p5-c0p01}
    \end{figure}

    In Fig.~\ref{fig:pcorr-k0p5-c0p01} we have plotted the static charge potential as a function of lattice separation.
    Here we take the continuum limit and keep the ratio $\beta / V = 0.01$ fixed while increasing the size of the lattice.  This figure is at a relatively small value of $\kappa = 0.5$, which according to Fig.~\ref{fig:kappa-sup-0p01}, puts this run well in the confining regime.  Here we see that the string breaking at small volumes is dominated by a linear potential in the continuum.
    
\begin{figure}[t]
        \centering
        \includegraphics[width=8.6cm]{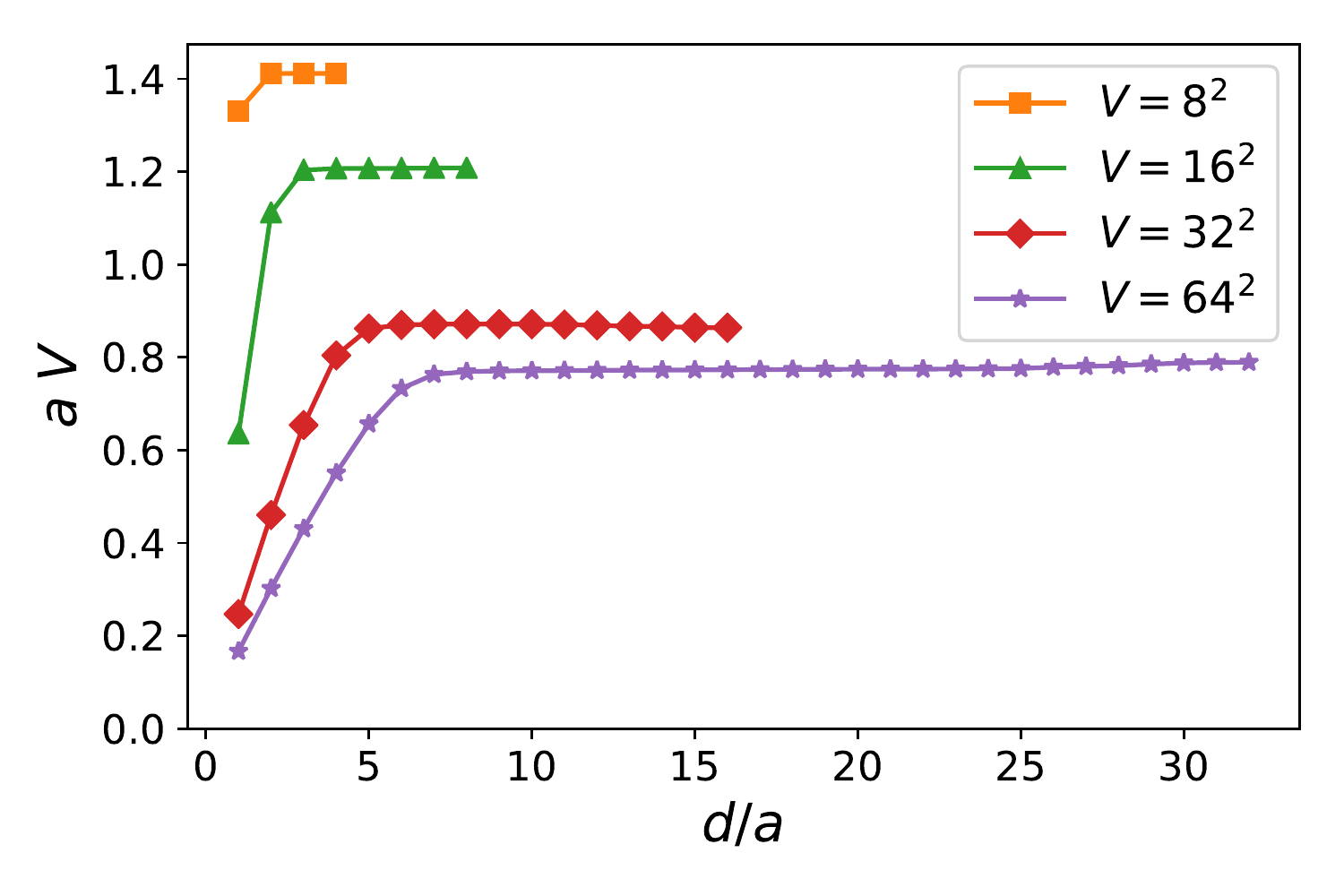}
        \caption{The static charge potential, $V$, from the Polyakov correlation function in the continuum limit.  Here we took $\beta / N_{s}N_{\tau} = 0.01$ fixed and varied the size of the lattice.  The gauge-matter coupling $\kappa$ is $2$, which is in the Higgs regime.  This is marked by string breaking at small distances between the static charges.}
        \label{fig:pcorr-k2p-c0p01}
    \end{figure}

    This is to be contrasted with Fig.~\ref{fig:pcorr-k2p-c0p01} where data was collected at a relatively larger $\kappa = 2$ value.  
    Again, Fig.~\ref{fig:kappa-sup-0p01} puts this kappa value in the Higgs regime.  A noticeable difference in the potential in Fig.~\ref{fig:pcorr-k2p-c0p01} is the much earlier onset of string breaking.  We notice that at small volumes, regardless of $\kappa$, string breaking occurs at short distances; however, as the continuum limit is approached the Higgs and confining regimes clearly separate.

\subsubsection{$\beta = 0$}

\begin{figure}[t]
    \centering
    \includegraphics[width=8.6cm]{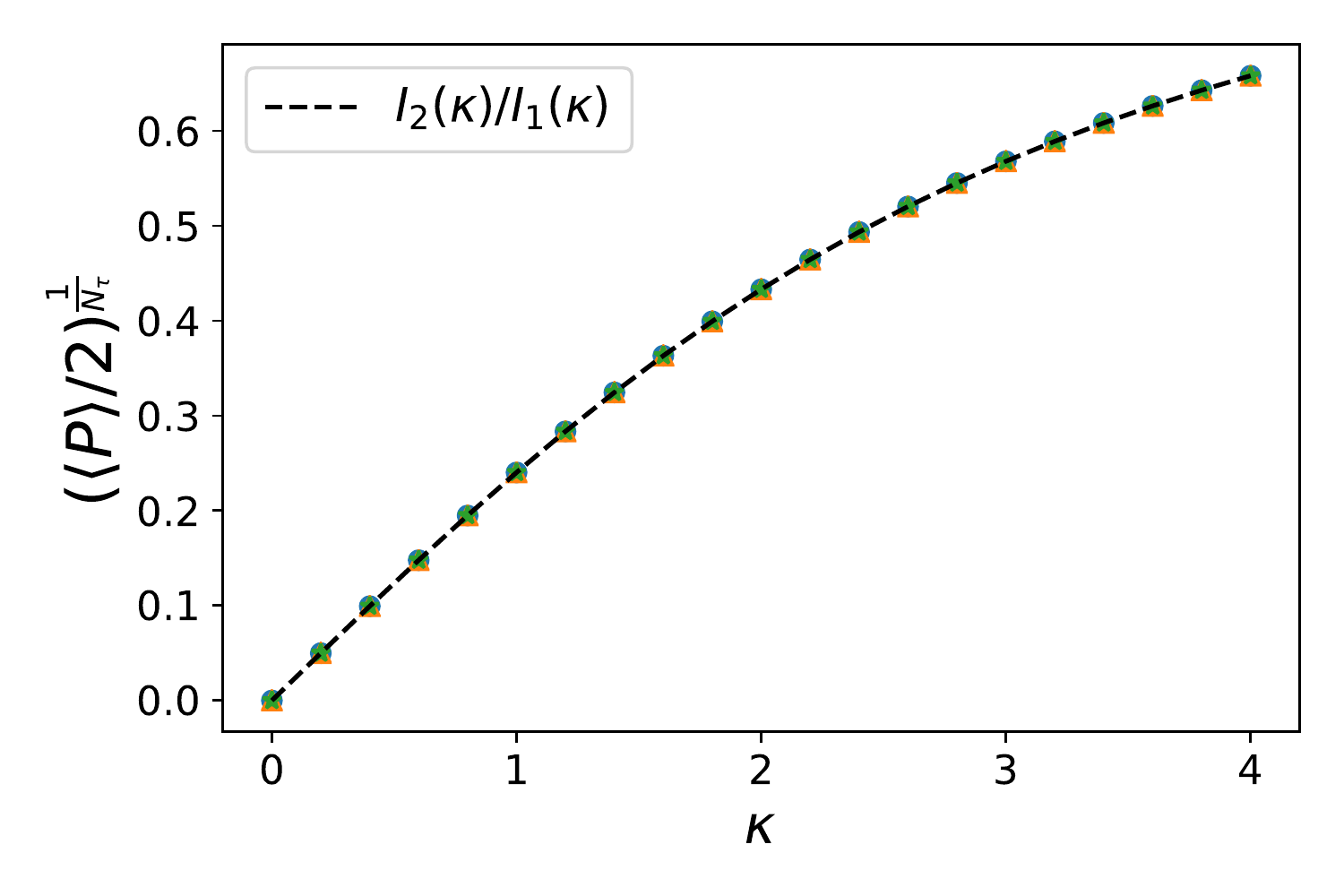}
    \caption{The Polyakov loop calculated using the TRG compared with the exact analytic calculation for the case of infinite gauge coupling ($\beta = 0$).  The $x$-axis is the gauge-matter coupling $\kappa$, and the $y$-axis is the re-scaled data compared with the exact dashed line from Eq.~\eqref{eq:beta0-ploop}, with $I_{1}$, and $I_{2}$ being the modified Bessel functions.  A few system sizes are plotted indicating a complete factorization of the Polyakov loop in this limit as expected.}
    \label{fig:exact-ploop-b0}
\end{figure}

    The $\beta = 0$ limit is trivial, but provides a simple check.  Looking at Eq.~\eqref{eq:btensor} we see that the only surviving representation on the lattice is the $r=0$ representation.  Equation~\eqref{eq:ploopA} then demands that only the $\sigma = 1/2$ contributes to the $\tilde{A}$ tensor.  The Clebsch-Gordan coefficients reduce to Kronecker deltas, and $\tilde{A}$ is diagonal in $i$ and $j$,
    \begin{equation}
        \tilde{A}^{(\tau)}_{0 0 i j}(\kappa) = \frac{1}{2}f_{\frac{1}{2}}(\kappa)\delta_{ij}.
    \end{equation}
    Then
    \begin{equation}
    \label{eq:beta0-ploop}
        \langle P \rangle = \frac{2}{Z} \left( \frac{f_{\frac{1}{2}}(\kappa)}{2}\right)^{N_{\tau}} = 2 \left(\frac{I_{2}(\kappa)}{I_{1}(\kappa)} \right)^{N_{\tau}},
    \end{equation}
    while,
    \begin{equation}
        G_{P P^{\dagger}} = 4 \left(\frac{I_{2}(\kappa)}{I_{1}(\kappa)} \right)^{2 N_{\tau}}.
    \end{equation}
A example of this collapse for the Polyakov loop can be seen in Fig.~\ref{fig:exact-ploop-b0} across a few of system sizes.
    
\subsubsection{$\kappa = 0$}
This limit is only slightly more complicated. In this case, for periodic boundary conditions, the Polyakov loop must vanish identically, as we would expect in the pure Yang-Mills theory.  This is because, on either side of the Polyakov loop, all the representation numbers for each plaquette must be identical, for the same reasons they were in Sec.~\ref{sec:kappa0}.  However, the Polaykov loop inserts a static charge of value $r=1/2$, and looking at Eq.~\eqref{eq:ploopA} we see that only the $\sigma=0$ representation survives, forcing either $r_l$, or $r_r$ to be incremented (or decremented) by $1/2$.  Since space closes in a circle, the representations cannot all be equal, and half be incremented by 1/2, simultaneously.  Therefore, the Polyakov loop vanishes.

The correlation function on the other hand remains finite.  This is because the Polyakov loop and its adjoint bound a region were all the constraints are satisfied, even with periodic boundary conditions.  In this limit the $\tilde{A}$ tensor associated with the Polyakov loop only allows for the $\sigma = 0$ representation in the sum.  This forces the representations on the plaquettes separated by the Polyakov loops to be shifted from each other by 1/2.  Similarly, the $m$ quantum numbers associated with the matrix indices are forced to conserve their $U(1)$ (or $O(2)$) charge across the Polyakov loop boundaries, which will be either $\pm 1/2$.  The Clebsch-Gordan coefficients reduce to Kronecker deltas.  Because $\kappa = 0$, similar to Sec.~\ref{sec:kappa0}, all the plaquette representation numbers in the bounded region must be the same, just as it must be outside the region as well.  However the two regions can differ by $\Delta r = 1/2$.  The correlation function can then be written as
\begin{align}
\nonumber
    &G(d, \beta) = \frac{1}{Z} \left[ \left( \frac{f_{0}(\beta)}{d_{0}} \right)^{(N_{s} - d)N_{\tau}} \left( \frac{f_{ \frac{1}{2}}(\beta)}{d_{ \frac{1}{2}}} \right)^{d N_{\tau}} \right. \\
    &\left. + \sum_{r = 1/2}^{\infty} \sum_{r'=r-\frac{1}{2}}^{r+\frac{1}{2}} \left( \frac{f_{r}(\beta)}{d_{r}} \right)^{(N_{s} - d)N_{\tau}} \left( \frac{f_{r'}(\beta)}{d_{r'}} \right)^{d N_{\tau}} \right]
\end{align}
for a separation of $d$ between the two Polyakov loops.

\section{Conclusions}
\label{sec:conclusions}
In this paper
we have derived a tensor formulation for the $SU(2)$ non-Abelian gauge-Higgs model. 
Using this tensor formulation, we have calculated observables using a renormalization group procedure known as the HOTRG, 
compared the results with Monte Carlo calculations and found good agreement. In addition, we have
studied the Polyakov loop and the correlation function between a Polyakov loop and its adjoint.  The
latter yields the static quark potential in the model.  While the system being two dimensional
can not exhibit a true phase transition the matter susceptibility, nevertheless, shows signs of a crossover
around $\kappa\sim 1.2$. The behavior of the correlation function suggests a confining regime for small $\kappa$ values, and 
a string breaking or Higgs-like phase for larger $\kappa$. We have also calculated the mass gap as a function of the matter-gauge coupling.
This is possible because the HOTRG method allows
for a straightforward construction of the transfer matrix by blocking only along a time slice.  From this approximate transfer matrix, the relative energy eigenvalues can be obtained. The crossover seen in the thermodynamics and static potential is also visible in the massgap which exhibits
different behaviors as $\kappa$ is varied.

One additional, important point to emphasize is that the TRG algorithm is able to calculate quantities like the static
quark potential that are essentially
impossible to calculate straightforwardly with Monte Carlo methods over large ranges of the parameter
space because of an exponentially small signal which is
swamped by noise. This problem is reminiscent of theories with a sign problem and indeed one of the principle advantages
of tensor network methods is their ability to completely avoid sign problems. 

Recasting the non-Abelian gauge-Higgs model in terms of the irreducible representations of the gauge group could allow a rotor formulation of the Hamiltonian of this model in the continuous-time limit.  There are already indications of the final form in 
Eq.~\eqref{eq:ym-ham} for the kinetic term, although further work is needed to describe the matter-gauge coupling term.  Such a formulation could provide a straightforward, gauge-invariant means to formulate the model for 
quantum simulations in the future.

\begin{acknowledgments}
The authors would like to thank Yannick Meurice for stimulating conversations connected with this work.
SC, RGJ, and JU were supported by the US Department of Energy (DOE), Office of Science, Office of High Energy Physics, under Award Numbers DE-SC0009998 and DE-SC0019139. AB was supported under grant DE-SC0019139.
\end{acknowledgments}



\bibliographystyle{utphys}

\end{document}